\documentclass[12pt]{article}
\usepackage{graphicx}
\usepackage{amsmath}
\usepackage{natbib}
\begin{document}

\begin{flushleft}
{\Large
\textbf{Bayesian Analysis of Epidemics - Zombies, Influenza, and other Diseases}
}
\\
Caitlyn Witkowski$^{1,\ast}$, 
Brian Blais$^{1,2}$
\\
\bf{1} Science and Technology Department, Bryant University, Smithfield RI 02917
\\
\bf{2} Institute for Brain and Neural Systems, Brown University, Providence RI
\\
$\ast$ Email: {\tt cwitkows@bryant.edu}
\end{flushleft}

\section*{Abstract}
Mathematical models of epidemic dynamics offer significant insight into predicting and controlling infectious diseases. The dynamics of a disease model generally follow a susceptible, infected, and recovered (SIR) model, with some standard modifications.  In this paper, we extend the work of Munz et.al  (2009)\nocite{munz2009zombies} on the application of disease dynamics to the so-called ``zombie apocalypse'', and then apply the identical methods to influenza dynamics.  Unlike Munz et.al  (2009), we include data taken from specific depictions of zombies in popular culture films and apply Markov Chain Monte Carlo (MCMC) methods on improved dynamical representations of the system.  To demonstrate the usefulness of this approach, beyond the entertaining example, we apply the identical methodology to Google Trend data on influenza to establish infection and recovery rates.  Finally, we discuss the use of the methods to explore hypothetical intervention policies regarding disease outbreaks.

\section{Introduction}
\label{introduction}

\begin{figure}
\begin{center}
\begin{tabular}{c|cc|} \cline{2-3}
\raisebox{.3in}{\textbf{\large A}} &\begin{minipage}{1.3in}
\begin{eqnarray*}
S'&=& -\beta SI \\
I'&=& +\beta SI -\zeta I \\
R'&=& +\zeta I\\
\end{eqnarray*}
\end{minipage}
&\raisebox{-.3in}{\includegraphics[height=0.6in]{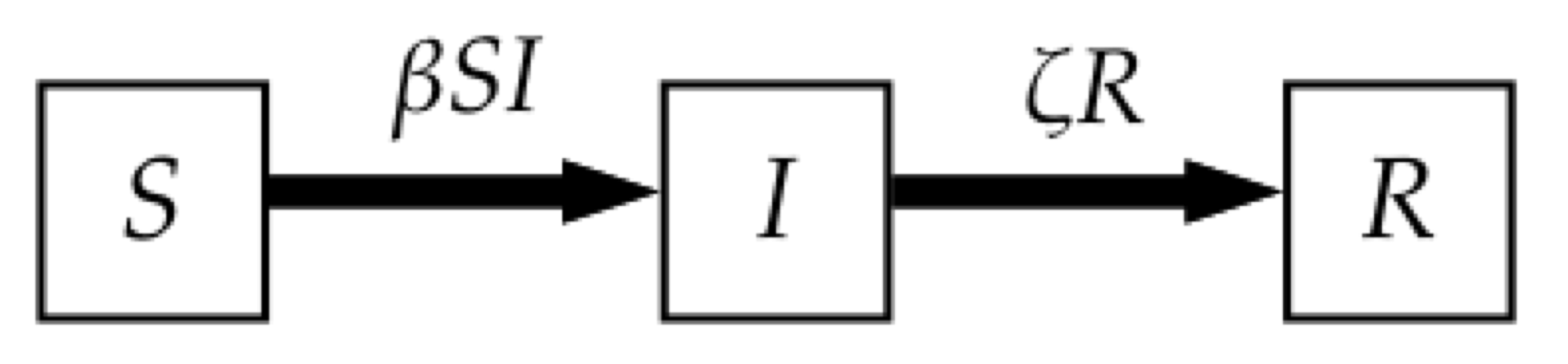}}\\\cline{2-3}
\raisebox{.3in}{\textbf{\large B}} &\begin{minipage}{1.3in}
\begin{eqnarray*}
S'&=& -\beta SI \\
E'&=& +\beta SI -\zeta E \\
I'&=& +\zeta E -\alpha I \\
R'&=& +\alpha I\\
\end{eqnarray*}
\end{minipage}
&\raisebox{-.3in}{\includegraphics[height=.57in]{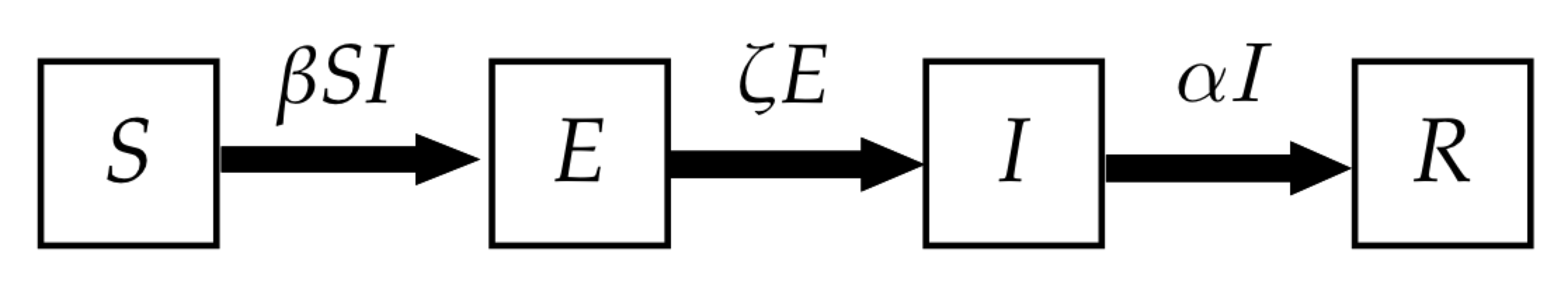}}\\\cline{2-3}
\raisebox{.3in}{\textbf{\large C}} &\begin{minipage}{1.5in}
\begin{eqnarray*}
S'&=& -\beta SZ \\
Z'&=& +\beta SZ +\zeta R -\alpha SZ \\
R'&=& +\alpha SZ -\zeta R \\
\end{eqnarray*}
\end{minipage}
&\raisebox{-.3in}{\includegraphics[height=0.75in]{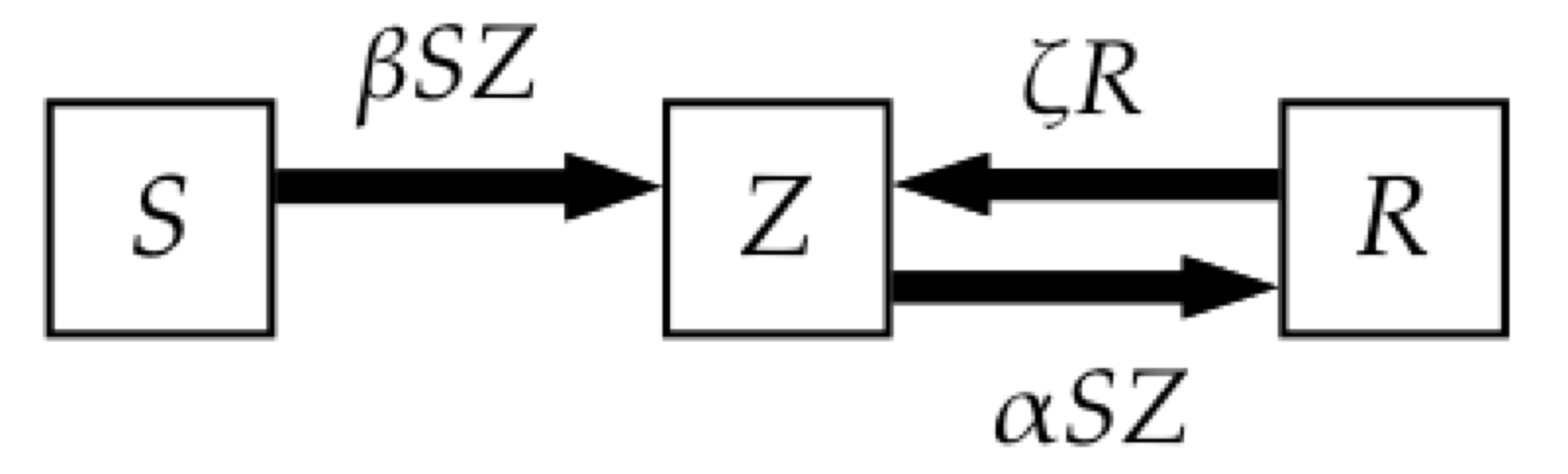}}\\\cline{2-3}
\raisebox{.3in}{\textbf{\large D}} &\begin{minipage}{1.3in}
\begin{eqnarray*}
S'&=& -\beta SZ -\delta S \\
E'&=& +\beta SZ - \zeta E \\
Z'&=& +\zeta E - \alpha SZ \\
R'&=& +\alpha SZ +\delta S \\
\end{eqnarray*}
\end{minipage}
&\raisebox{-.3in}{\includegraphics[height=0.8in]{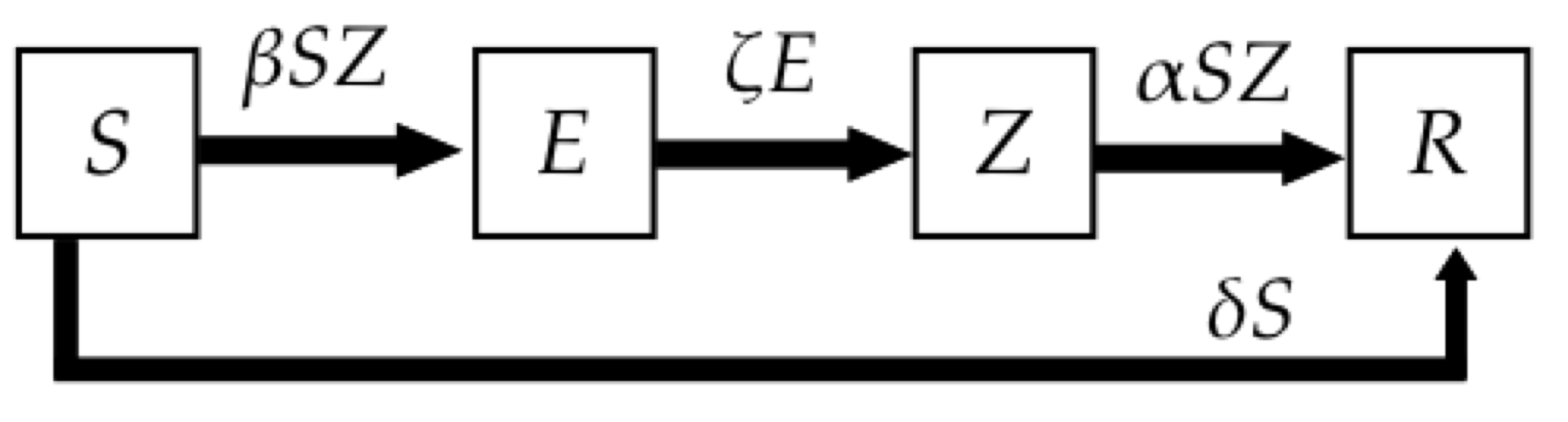}}\\\cline{2-3}
\raisebox{.3in}{\textbf{\large E}} &\begin{minipage}{1.3in}
\begin{eqnarray*}
S'&=& -\beta SZ \\
E'&=& +\beta SZ -\zeta E \\
Z'&=& +\zeta E -\alpha SZ \\
R'&=& +\alpha SZ \\
\end{eqnarray*}
\end{minipage}
&\raisebox{-.3in}{\includegraphics[height=0.7in]{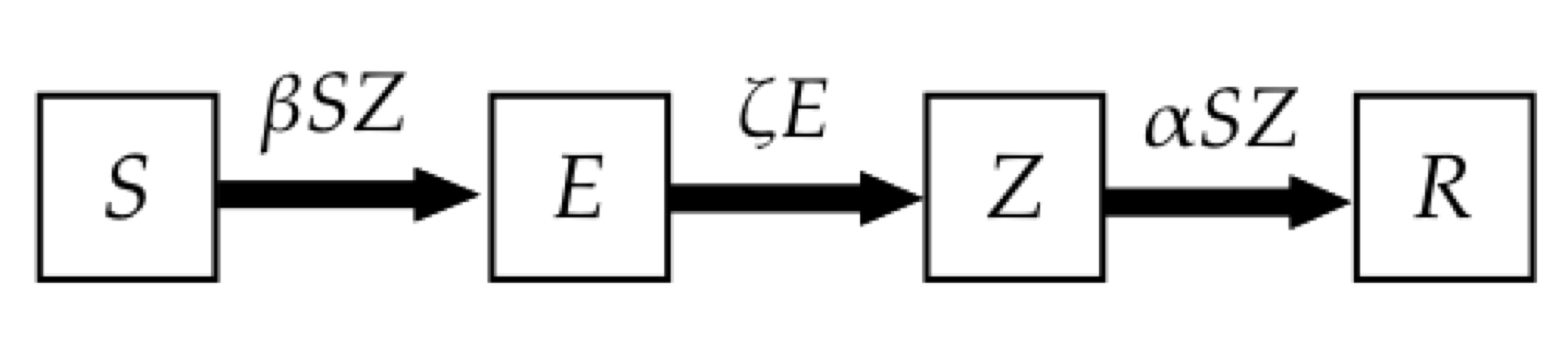}}\\\cline{2-3}
\end{tabular}
\end{center}
\caption{Systems diagrams for models of disease and zombie dynamics.  Shown are (A) the standard SIR model, (B) the standard SEIR model, (C) the Munz et. al. (2009) model, (D) the model for \emph{Night of the Living Dead}, and (E) the model for \emph{Shaun of the Dead}.}
\label{fig:block_diagrams}
\end{figure}

Epidemic models have combined mathematical modeling and biological system dynamics to predict and control infectious diseases, as seen in studies on measles \citep{mcgilchrist1996loglinear,grais2006estimating,tuckwell2007some,kuniya2011global} and influenza \citep{tuckwell2007some,li2009stability,hooten2010assessing,coelho2011bayesian}, for example. Commonly referred to as SIR models, these dynamical epidemic models describe disease dynamics via subgroups susceptible (S), infected (I), and recovered (R) with parameters related to rate of infectiousness and recovery, as seen in Figure~\ref{fig:block_diagrams}A these parameters can be modified to different infectious disease dynamics. 

The dynamical equations for the standard SIR model\citep{Kermack:1927fk} are:

\begin{equation}\label{eq:SIR}
\begin{split}
S'&= -\beta SI \\
I'&= +\beta SI -\zeta I \\
R'&= +\zeta I
\end{split}
\end{equation}

A typical extension is what is referred to as the SEIR model\citep{aron1984seasonality}, where the fourth component ($E$ for exposed) represents an infected but not symptomatic or infectious population (see Figure~\ref{fig:block_diagrams}).  
\begin{equation}\label{eq:SEIR}
\begin{split}
S'&= -\beta SI \\
E'&= +\beta SI -\zeta E \\
I'&= +\zeta E -\alpha I \\
R'&= +\alpha I
\end{split}
\end{equation}

Although SIR-type mathematical models have a long history of being applied to biological systems, Bayesian parameter estimation applied to dynamical models is a relatively new but critical addition. Using these methods, a statistical model is developed for the parameters of the \emph{dynamical} model. Markov Chain Monte Carlo (MCMC) techniques are then applied to estimate the posterior probabilities of the parameters, providing both the best estimates and their uncertainty~\citep{coelho2011bayesian}. 

This work attempts to use an entertaining example, that of the impending so-called zombie apocalypse, to demonstrate both the methods of model construction and the Bayesian analysis of statistical models applied to dynamical systems. The reason for using this particular example, as opposed to a more traditional disease like influenza, is that entertaining examples can improve student learning of a topic, makes a topic more accessible to an outside audience and thus to the public understanding of science\citep{ziman1991public}, and is fun way to explore a new technical tool for researchers. All of the specific tools and methods examined here can be applied to more traditional diseases, which we demonstrate later in the case of influenza (Section~\ref{sec:discussion}). The idea of applying epidemic modeling to an outbreak of zombie infection is credited to Munz et. al. (2009)~\nocite{munz2009zombies}.  The current paper corrects some inaccuracies in this original work, applies the tools of Bayesian parameter estimation to the problem, and (in contrast to Munz et.al. (2009)) uses quantitative data to estimate the model parameters - thus making the methods directly applicable to real-world disease dynamics.

\subsection{Previous Work}
\label{previouswork}

The term ``zombie'' either originated from the North Mbundu term `nzumbe' in Africa or from the Haitian Creole term `zonbi' in the Caribbean, both of which refer to a corpse reanimated via witchcraft\citep{dictionary2010oxford}. In modern usage, zombie still refers to the reanimated dead, however the specifics of the term have changed based on depictions in films and television shows. Some movies still portray the voodoo zombie\citep{bravman1986,steckler1964} but almost all pop culture zombies resemble disease dynamics\citep{romero1968,romero1978,romero1985,wright2004,fleischer2009,russo1985,anderson2002,anderson2004,anderson2007,anderson2010,anderson2012,darabont2010,boyle2002,boyle2007,evjen2009,clark1972,schenkman2012}. Although zombies differ from the typical infected person in an epidemic model (hunger for human flesh and difficulty to kill), the dynamics are remarkably similar; the zombie dynamics is well described with susceptible and infected subpopulations, with transitions between these model states. 

The idea of applying epidemic modeling to an outbreak of zombie infection is first presented in Munz et. al. (2009)\nocite{munz2009zombies}. They apply an SIR model to the entertaining example of a zombie attack, then continue to determine equilibria and stability of the models under a number of conditions.  Although Munz et. al. (2009) claims to apply the model to popular movie depictions of zombies, to our knowledge the dynamics exhibited in their paper are unlike \emph{any} in pop culture media: neither film, television, comic book, nor video game. Figure~\ref{fig:block_diagrams}C shows the basic model proposed by Munz et. al. (2009), where the resulting dynamical equations are:

\begin{equation}\label{eq:Munz}
\begin{split}
S'&= -\beta SZ \\
Z'&= +\beta SZ +\zeta R -\alpha SZ \\
R'&= +\alpha SZ -\zeta R 
\end{split}
\end{equation}
Here $S$, $Z$, and $R$ are the susceptible, zombie (i.e. infected), and removed (i.e. recovered) subpopulations, respectively. In the SIR and SEIR models, the R term doesn't represent dead, but simply removed from the susceptible population.  In the zombie models, the only way to remove someone from the susceptible population is through death, but the term serves the same function mathematically in all models.  The parameter $\beta$ quantifies the infection rate, $\alpha$ the rate of removal (i.e. killing) of the zombies through an interaction with a susceptible, and $\zeta$ the rate of return of the zombies from the dead. 

It is this last parameter, and the resulting term $\zeta R$, which is unlike any film.  One serious consequence of this term is that there is no \emph{permanently} removed subpopulation, suggesting that in their model the zombies never truly die.  Although films portray the vast difficulty in destroying a zombie, in every movie depiction we've seen, it is always possible to permanently remove a zombie (e.g. cut off the head, destroy the brain, burn the body, etc.). This basic model, which we refer to as the Munz et. al. (2009) model (Figure~\ref{fig:block_diagrams}C), is the basis for all of the models in the entire Munz et al. (2009) paper. The assumption that the removed population will inevitably recycle into the zombie subpopulation implies that the zombie subpopulation will always win, while the other two subpopulations will always be diminished, regardless of parameters. We have found that this assumption does not match \emph{any} depiction of zombies in the popular culture, so the conclusions from models with this assumption should be suspect.  Further, there is no analog to this ``recyling'' terms in real-world diseases, thus limiting the application of the Munz et. al. (2009)  model to only entertainment purposes.

\section{Methods}
\label{methods}

\subsection{The Films and Model Structure}
\label{thefilmsandmodelstructure}

To gain insight into zombie dynamics, we examine a large variety of zombie films\citep{bravman1986,steckler1964,romero1968,romero1978,romero1985,wright2004,fleischer2009,russo1985,anderson2002,anderson2004,anderson2007,anderson2010,anderson2012,darabont2010,boyle2002,boyle2007,evjen2009,clark1972,schenkman2012}. From these films and television shows, we find that virtually all zombie movies fall into one of two forms of the dynamical epidemic model, shown in Figure~\ref{fig:block_diagrams}D and Figure~\ref{fig:block_diagrams}E. We fit each of these forms to a specific case study, the first model represented by \emph{Night of the Living Dead}~\citep{romero1968} and the second by \emph{Shaun of the Dead}~\citep{wright2004}, exploring each in turn.

\subsubsection{\emph{Night of the Living Dead}}
\label{nightofthelivingdead}

Considered the first depiction of zombies in the popular culture, Romeros' three films\citep{romero1968,romero1978,romero1985} (\emph{Night of the Living Dead}, \emph{Dawn of the Dead}, and \emph{Day of the Dead}) all take place in the same apocalyptic world and thus exhibit the same dynamics, as shown in Figure~\ref{fig:block_diagrams}D. The dynamical equations are derived from the following observations from the movies:

\begin{enumerate}
\item Anyone who dies becomes a zombie, regardless of contact with one.

\item Because contact with a zombie is likely to lead to death, the interaction between the two subpopulations of susceptibles and zombies is significant.

\item This interaction between susceptibles and zombies results in a temporarily removed subpopulation before members of that population become zombies.

\item The only way in which a zombie can be permanently removed is by destroying the brain or burning the body.

\end{enumerate}

These observations yield the following dynamical equations:
\begin{equation}\label{eq:NightLivingDead}
\begin{split}
S'&= -\beta SZ -\delta S \\
E'&= +\beta SZ - \zeta E \\
Z'&= +\zeta E - \alpha SZ \\
R'&= +\alpha SZ +\delta S 
\end{split}
\end{equation}
where, like the standard SEIR model (Equation~\ref{eq:SEIR}), the interaction parameter $\beta$ quantifies the rate at which the susceptible human subpopulation ($S$) moves into the exposed population ($E$) through interaction with a zombie ($Z$).  The exposed ($E$) term is a temporarily removed subpopulation (i.e. removed from the susceptible population) in all models, although the mechanism is different in each.  For example, in SEIR models it represents those people who have the disease but are not contagious while in \emph{Night of the Living Dead} it is anyone who has died - and thus will become a zombie.  Despite these differences, this term serves the same function mathematically in all models.

The temporarily removed subpopulation ($E$) then automatically moves into the zombie subpopulation ($Z$) at a fixed rate $\zeta$ in relation to the size of the exposed subpopulation size. The interaction term $\alpha SZ$ represents the susceptible permanently killing a zombie, as described above.  The most significant difference between this model and the typical SEIR model is that in the zombie model one requires an interaction between a susceptible and ``infected'' to move the ``infected'' into the removed subpopulation (i.e. an uninfected person must kill a zombie to remove them), whereas in the SEIR model, infected individuals recover without any interaction.  This difference leads to a more difficult ``infection'' to eliminate in the zombie model compared to the SEIR model.

Although zombies can starve, it takes months for this to happen and the zombie apocalypse takes days, so a term for starvation is neglected in the model. Finally, $\delta S$ represents susceptibles who commit suicide; despite the rarity of suicide in a typical human population, in the Romero films, a rather large percent of susceptibles killed themselves to avoid being eaten alive and/or entering the zombie subpopulation. 

\subsubsection{\emph{Shaun of the Dead}}
\label{sec:shaunofthedead}

Although the film's title is a play onâ \emph{Dawn of the Dead}, the zombie farceâ \emph{Shaun of the Dead}~\citep{wright2004} has a very different dynamical structure, as shown in Figure~\ref{fig:block_diagrams}E. We use \emph{Shaun of the Dead} because it is the most typical representation of popular zombie films. The greatest difference between this model to the previous model proposed is that contact between a susceptible and zombie is necessary for the zombie population to grow - not all susceptibles who die become zombies. 

\begin{equation}\label{eq:ShaunDead}
\begin{split}
S'&= -\beta SZ \\
E'&= +\beta SZ -\zeta E \\
Z'&= +\zeta E -\alpha SZ \\
R'&= +\alpha SZ 
\end{split}
\end{equation}

The specifics of the zombie dynamics may vary slightly from film to film, but in general, the interaction parameter $\beta$ has the same effect of quantifying the infection rate, moves the susceptible subpopulation ($S$) into the exposed subpopulation ($E$).  The exposed subpopulation moves automatically into the zombie subpopulation ($Z$) at a rate of $\zeta E$. Finally, as seen in the previous model proposed, zombies can move into a \emph{permanently} removed subpopulation ($R$) via an interaction with susceptibles, $\alpha SZ$, in which a susceptible destroys a zombie's brain. Suicide rates are ignored in this model because there were no suicides in \emph{Shaun of the Dead}, however a suicide term from ($S$) to ($R$) could be added if one were examining the dynamics of a different zombie world, such as the popular television showâ \emph{Walking Dead}, where suicide is commonplace.

\subsubsection{Simplified Form}

If $\delta$ is small in Equation~\ref{eq:NightLivingDead} (i.e. suicides are neglected), and $\zeta$ is large  in Equations~\ref{eq:NightLivingDead} and~\ref{eq:ShaunDead} (i.e. exposed people become zombies within minutes), then both models reduce to the following simplified form:
\begin{equation}\label{eq:ShaunDeadSimple}
\begin{split}
S'&= -\beta SZ \\
Z'&= +\beta SZ -\alpha SZ \\
R'&= +\alpha SZ 
\end{split}
\end{equation}

In this paper we work with the full form for each of the models, but the connection with the standard SIR model (Equation~\ref{eq:SIR}) becomes more clear in this approximation.

\subsection{Data Collection}
\label{datacollection}

Although real data on zombies is non-existent, we collected approximate population estimates from the films themselves at various time-points. Zombie numbers were estimated in scenes with a field of view approximately 50 meter squared area, and the time values estimated from visual cues (clocks, sun, estimated time characters completed tasks, etc{\ldots}). Although this process almost certainly contains large uncertainties, the process for estimating the number of people infected with influenza is plagued with similar uncertainties\citep{hooten2010assessing}. Table~\ref{tbl:movie_data} shows the data for both movies.

\begin{table}

\begin{center}
\begin{tabular}{ccc}

  \begin{tabular}{@{} cc @{}}
\multicolumn{2}{c}{\bf Shaun of the Dead} \\
    \hline
    Time (hours) & Zombies \\ 
    \hline
0.0& 0 \\
3.0& 1 \\
5.0& 2 \\
6.0& 2 \\
8.0& 3 \\
10.0& 3 \\
22.0& 4 \\
22.2& 6 \\
22.5& 2 \\
24.0& 3 \\
25.5& 5 \\
26.0& 12 \\
26.5& 15 \\
27.5& 25 \\
27.75& 37 \\
28.5& 25 \\
29.0& 65 \\
29.5& 80 \\
31.5& 100 \\
    \hline
  \end{tabular}
 &\hspace*{.3in}
    &
  \begin{tabular}{@{} cc @{}}
\multicolumn{2}{c}{\bf Night of the Living Dead} \\
    \hline
    Time (hours) & Zombies \\ 
    \hline
0.0& 1 \\
1.0& 1\\
1.5& 3\\
3.0& 8\\
4.5& 10\\
5.0& 20\\
5.75& 28\\
5.9& 30\\
10.0& 40\\
    \hline
  \end{tabular}
  \end{tabular}

\end{center}

\caption{Individual infected counts taken from movie depictions of zombies.}\label{tbl:movie_data}
\end{table}

\subsection{Estimating initial susceptible population}
\label{estimatinginitialparameters}

We can estimate the initial susceptible population from the movie data using a simple approximation.  If we assume that the zombies grow unrestricted (i.e. $\alpha\ll\beta$) and the population of susceptibles is initially large enough to not be significantly affected (i.e. $S\sim {\rm constant}\equiv S_{o}$), we can approximate the dynamics of zombies as exponential:

\begin{eqnarray}
Z'&=&(\beta S_{o}) Z
\end{eqnarray}
which has the solution

\begin{eqnarray}
Z&\sim& Z_{o}e^{\beta S_o t}
\end{eqnarray}
or a slope of $\beta S_{o}$ on a $\log Z$ plot.  We then can directly estimate $S_{o}$ from the data shown in Table~\ref{tbl:movie_data} for each movie.  

\subsection{Applying Bayesian Analysis Through MCMC}
\label{applyingbayesiananalysisthroughmcmc}

A statistical model of the parameters, $\beta$, $\alpha$, $\zeta$, and $\delta$, as well as the initial values for the populations ($S_o$, $E_o$, and $Z_o$), is attached to the dynamical model most appropriate for the given movie. Uniform prior probabilities are assumed for the parameters and the statistical Markov Chain Monte Carlo (MCMC) simulation\citep{patil2010pymc,gelman2004bayesian} is run between 100000-500000 iterations with a burn-in of 10\%. Following the MCMC iterations, histograms of the parameter traces are used to estimate the posterior probabilities for the parameter values, the uncertainty in those values, and the correlation between parameters in the models.

\subsection{Software}

The full computer code, written in Python, for both the Munz et. al. (2009)  models and our models, described presently, is included in the supplemental materials.  The dynamical models were simulated using a freely available wrapper around the {\tt odeint} function in Scipy, and the statistical models were constructed  using a wrapper around the PyMC numerical library\citep{patil2010pymc}.  These new tools allow one to do a sophisticated statistical analysis on a complex dynamical system with a minimal of technical syntax.

\section{Results}
\label{results}

\subsection{Stability}
\label{stability}

To solve for stability of the system around the point where $Z=0$, Munz et. al. (2009)\nocite{munz2009zombies} calculate the eigenvalues of the Jacobian,
\begin{eqnarray}
-\lambda \left(\lambda^{2}+\left(\zeta - (\beta-\alpha)N\right)\lambda-\beta\zeta N\right)=0
\end{eqnarray}
which has solutions
\begin{eqnarray}
\lambda&=&0 \\
\lambda_{+/-}&=&\frac{N(\beta-\alpha)-\zeta\pm\sqrt{(\zeta-(\beta-\alpha)N)^{2}+4\beta\zeta N}}{2}
\end{eqnarray}
which yields positive $\lambda$ for any value of the ``\emph{recycling}'' parameter, $\zeta$ - thus the zombies will win in any scenario. 

However, without the ``recycling'' parameter, $\zeta$ as used in Munz et.al. (2009), the stability is changed significantly.  If we remove this the ``recycling'' parameter, as we do in our models, here setting $\zeta=0$, we then obtain

\begin{eqnarray}
\lambda_{+/-}&=&\frac{N(\beta-\alpha)\pm\sqrt{((\beta-\alpha)N)^2}}{2}\\
\lambda_{-}&=&0\\
\lambda_{+}&=&2N(\beta-\alpha)
\end{eqnarray}
which results in $\lambda_{+}<0$ if $\alpha>\beta$. In other words, the zombies \emph{can be} completely removed, as long as we remove the zombies faster than they are growing. Thus the no-win situation described by Munz et. al. (2009)\nocite{munz2009zombies} is a direct by-product of their (incorrect) choice of model structure.

\subsection{Parameter Distributions}
\label{parameterdistributions}

\begin{figure}[htbp]
\begin{center}
\includegraphics[width=7in]{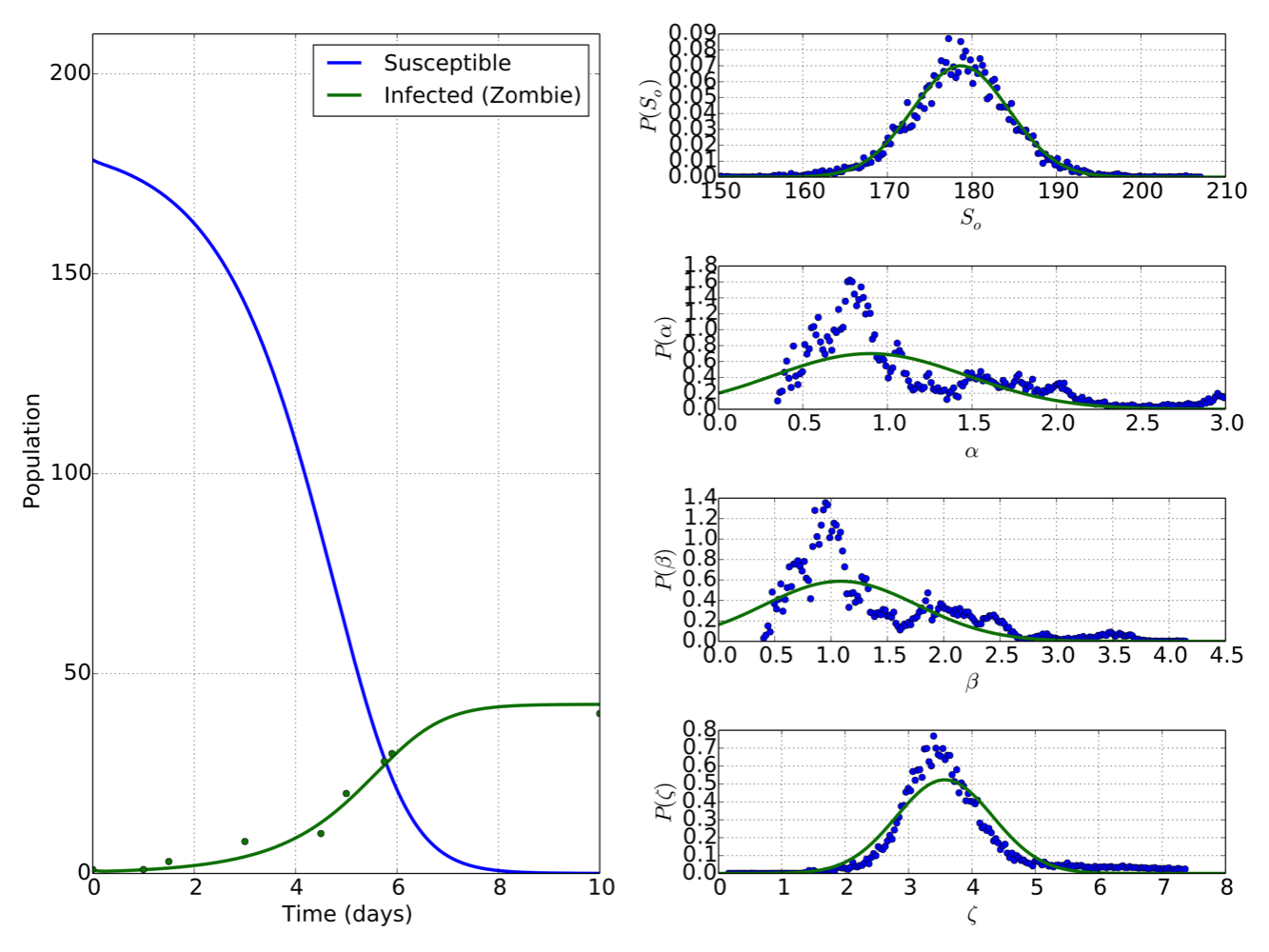}
\caption{The best-fit simulation (left) for the \emph{Night of the Living Dead} and the posterior probability distributions (right) for the parameters $S_{o}$, $\alpha$, $\beta$, and $\zeta$.}
\label{fig:nightparameters}
\end{center}
\end{figure}
\begin{figure}[htbp]
\begin{center}
\includegraphics[width=7in]{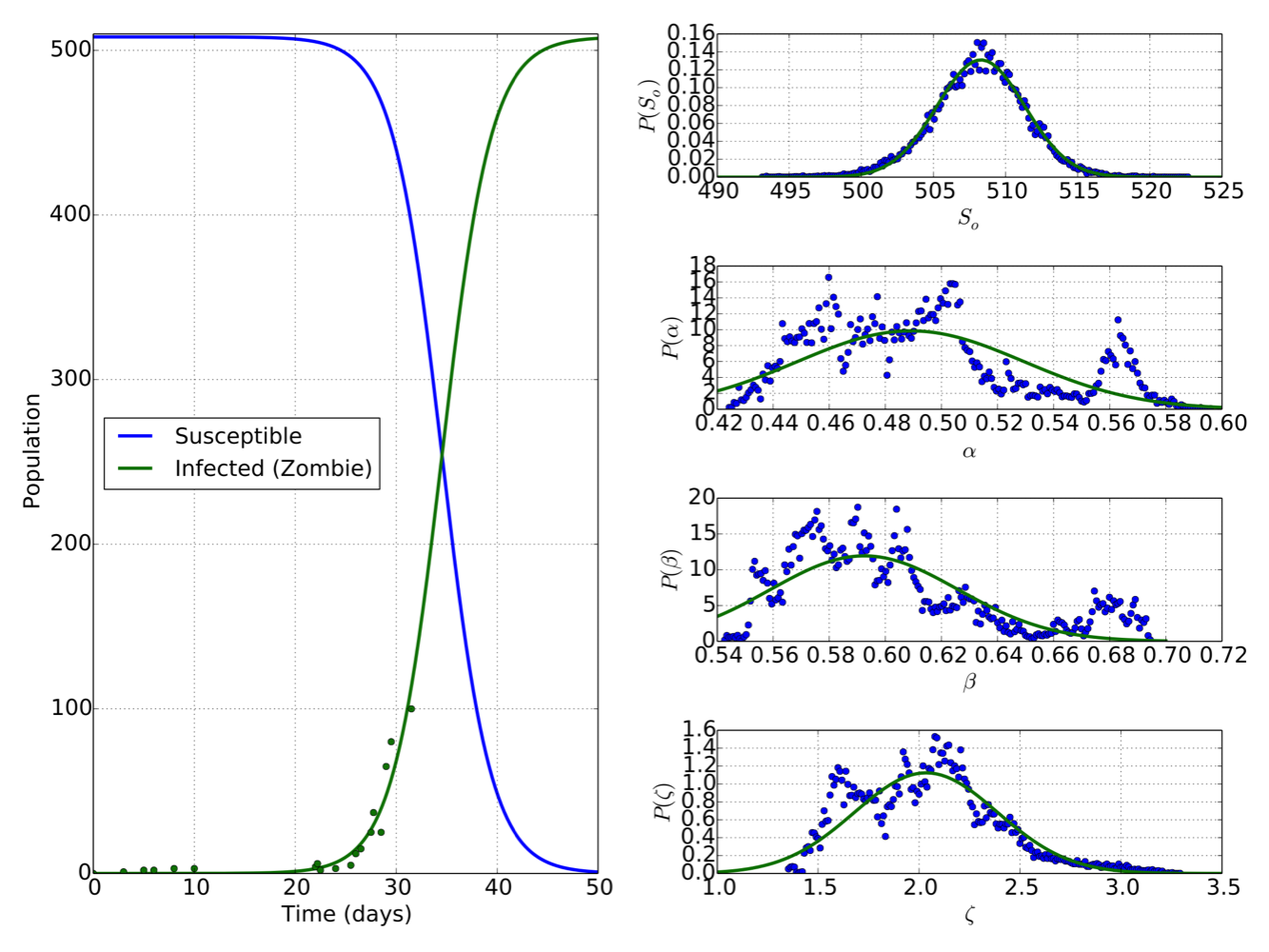}
\caption{The best-fit simulation (left) for the \emph{Shaun of the Dead} and the posterior probability distributions (right) for the parameters $S_{o}$, $\alpha$, $\beta$, and $\zeta$.}
\label{fig:shaunparameters}
\end{center}
\end{figure}
\begin{figure}[htbp]
\begin{center}
\begin{tabular}{cc}
\includegraphics[width=3.1in]{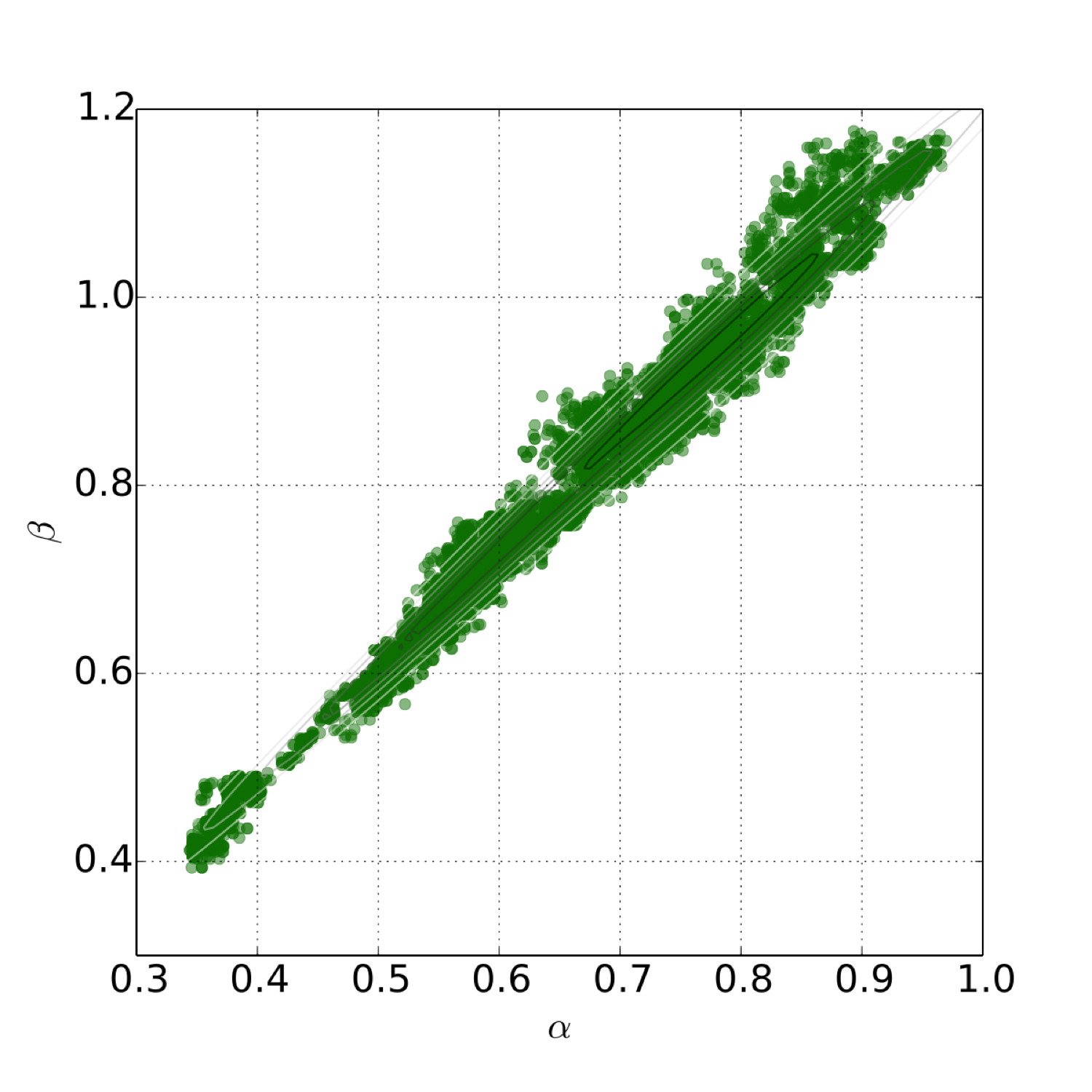}&
\includegraphics[width=3.1in]{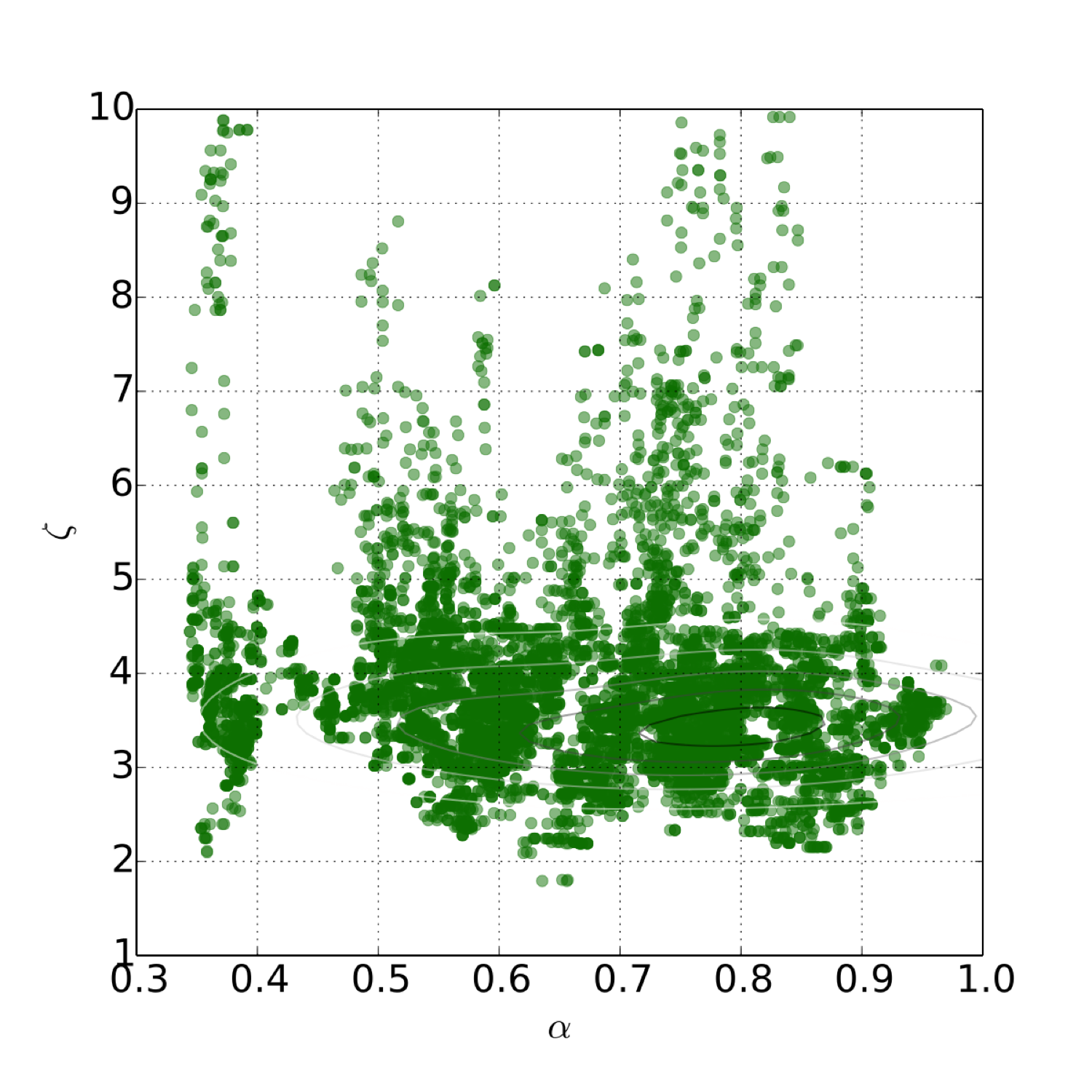}
\end{tabular}
\caption{The joint distribution between parameters $\alpha$ and $\beta$ (left) and $\alpha$ and $\zeta$ (right) for \emph{Night of the Living Dead}.  A clear relationship between $\alpha$ and $\beta$ is seen, where higher values of the infection rate ($\beta$) require a higher rate of removal of infection ($\alpha$) to fit the data well.  Conversely, no relationship can be scene between the rate of removal of infection ($\alpha$) and the rate of exposed becoming fully infected ($\zeta$)}
\label{fig:living_joint}
\end{center}
\end{figure}

Shown in Figures~\ref{fig:nightparameters} and ~\ref{fig:shaunparameters} are the parameter distributions from the dynamical models for \emph{Night of the Living Dead} and \emph{Shaun of the Dead}, respectively.  The summary statistics  are shown in Table~\ref{tbl:summary_parameters}.  Although the rates of infection and removal ($\beta$ and $\alpha$, respectively) are nearly a factor of two smaller for \emph{Shaun of the Dead} than \emph{Night of the Living Dead}, we can make a case for consistency by observing the joint distribution between these two variables, shown in Figure~\ref{fig:living_joint} for \emph{Night of the Living Dead}.  Here one can see that the parameters for both movies falls within this joint distribution.

The two movies, however, do seem differ in their values for the rate of exposed becoming fully infected ($\zeta$).  The joint distribution between $\zeta$ and $\alpha$, shown in Figure~\ref{fig:living_joint}, demonstrates no strong relationship between these two parameters consistent with the data.

\begin{table}[htdp]
\begin{center}
  \begin{tabular}{@{} cccc @{}}
    \hline
    & &  & Google Trends Influenza\\ 
    & Night of the Living Dead & Shaun of the Dead & Argentina 2011-2012\\ 
    Name & $\mu$ (95\% interval) &$\mu$ (95\% interval) &$\mu$ (95\% interval)   \\ 
    \hline
$S_o$ & 178.66 (166.03,190.62) &508.29 (501.58,514.26) &10000.35 (9938.41,10062.46) \\
$\alpha$ & 0.9 (0.42,2.7) &0.49 (0.44,0.57) &0.053 (0.051,0.055) \\
$\beta$ & 1.1 (0.51,3.3) &0.59 (0.55,0.69) &0.061 (0.059,0.065) \\
$\zeta$ & 3.6 (2.4,8.1) &2 (1.5,2.8) &N/A \\
    \hline
  \end{tabular}
\end{center}
\caption{Summary of the fit parameters.}
\label{tbl:summary_parameters}
\end{table}%

\section{Discussion}
\label{sec:discussion}

\begin{figure}[htbp]
\begin{center}
\includegraphics[width=7in]{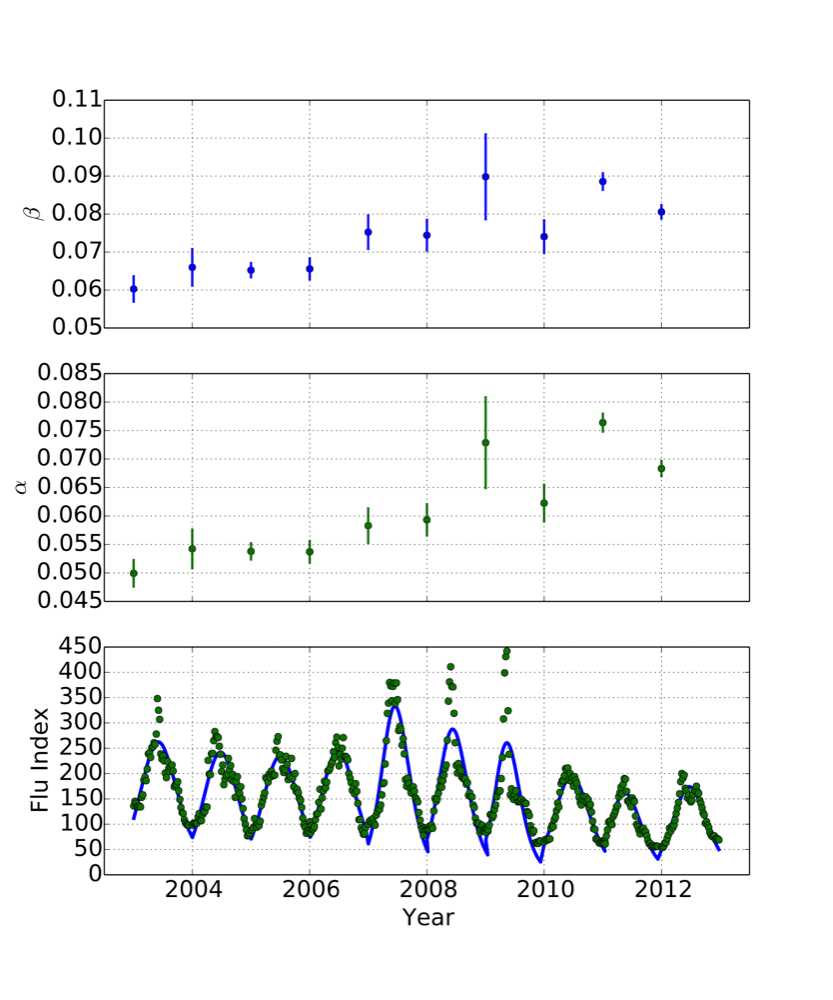}
\caption{The best-fit parameters for $\beta$ (top) and $\alpha$ (middle), for each Argentina influenza season, data obtained from Google Trends.  Error bars show the 95\% credible regions for the parameters. Shown is the best-fit time series (bottom) through several seasons of influenza.  Credible regions for the time series are also available, but for clarity are not shown.}
\label{fig:google_trend}
\end{center}
\end{figure}

The models presented here are a significant improvement over the models in Munz et. al. (2009)\nocite{munz2009zombies}, both in terms of model structure more closely matching the system of study (i.e. popular culture depictions of zombies) and the use of data collected from those systems. Using Bayesian parameter estimation, we've obtained best-fit values for the parameters in the dynamical model and quantified their uncertainty. Using \emph{exactly} the same techniques, we can analyze real-world data on influenza, as shown in Figure~\ref{fig:google_trend} where we've used Google trend data on influenza and the standard SIR model. The posterior probability distributions for the model parameters are particularly clear and consistent in this case, giving narrow error bars and a consistent time dynamics through several seasons.  It should be reiterated that this is the goal of this work - to present a straightforward method for handling real-world disease dynamics, motivated by an entertaining example.  

\begin{figure}[htbp]
\begin{center}
\includegraphics[width=7in]{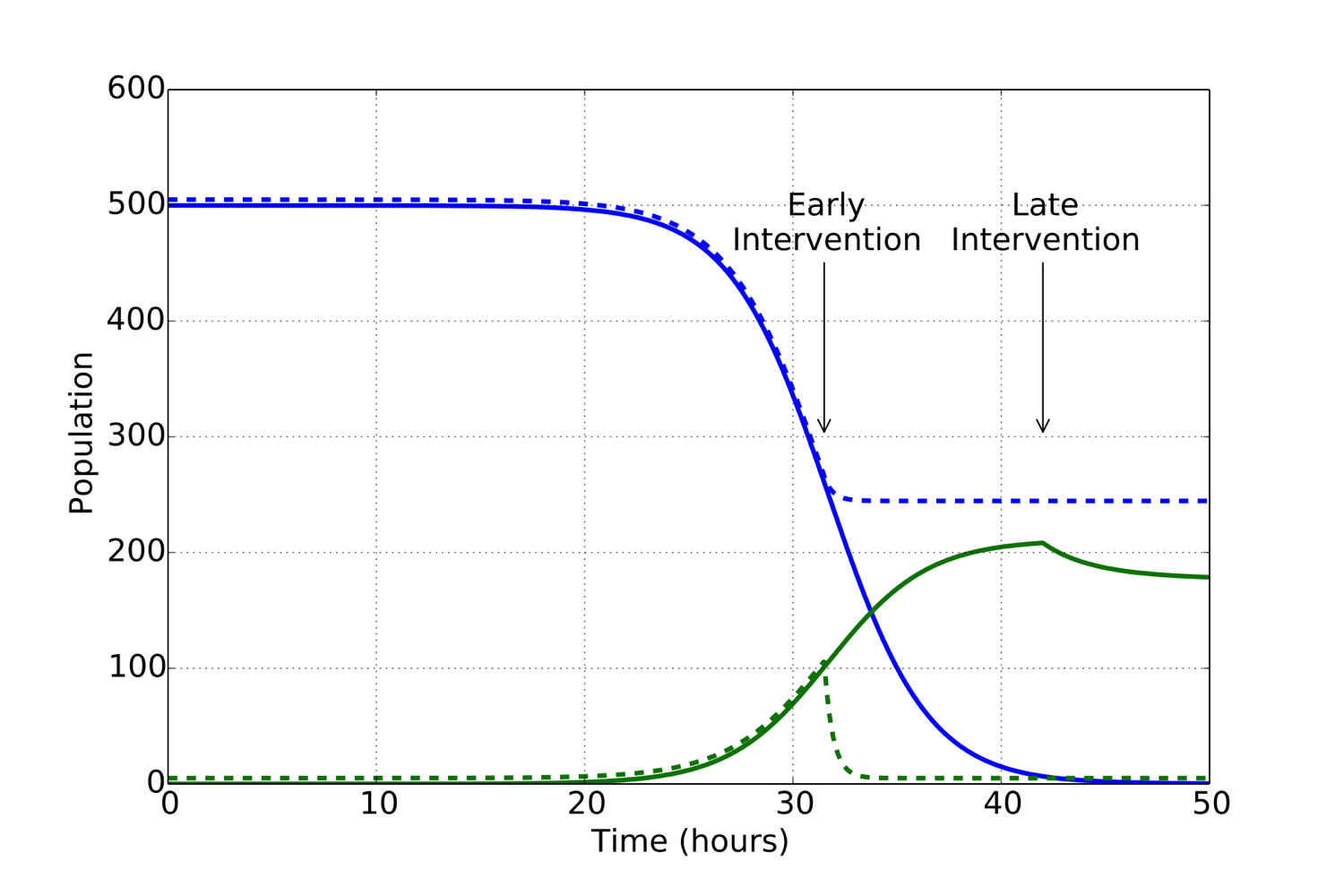}
\caption{The effect of the timing of military intervention.  Shown is the effect on the susceptible and zombie populations depending on a early or late military intervention.  For the model parameters where the zombies overtake the susceptible population, there is a clear sensitivity to the timing of intervention.}
\label{fig:intervention}
\end{center}
\end{figure}

Beyond best-fit estimates, we can also analyze hypothetical scenarios surrounding the system of study. For example, at the end of both films analyzed in detail here  a military intervention saves human civilization. The strength of the military attack and the time at which the intervention occurs are the two major factors in determining the success of this tactic. If the intervention happens quickly, the zombie population is quickly and expectedly annihilated; however, if the intervention happens later, intervention of any magnitude may prove fruitless. To explore this, inâ ``Shaun of the Dead'' we assume that the military intervention at the end of the film increased the $\alpha$ by ten-fold, effectively eliminating the zombie population and restoring civilization. However, if the film had lasted only half an hour longer, that same intervention strength would have led to an apocalyptic ending (shown in Figure~\ref{fig:intervention}).  Other hypothetical scenarios can also be explored in a straightforward way.

Studying different types of zombie dynamics, although obviously fictional, are extremely useful in constructing real-life disease models. We have demonstrated that the same model structure can be used in both cases, but that the zombie models have the added value of also being entertaining - a value which can aid in both education and outreach. Similarly, the same mathematical and statistical techniques are useful in both types of problems. It therefore becomes critical in avoiding the zombie epidemic, or any other epidemic, to have a sound understanding stemming of the dynamics of the disease from which to make forecasts. This understanding comes from constructing reasonable models applied to data on the initial spread of the disease, and applying reliable inference techniques such as those from Bayesian analysis.

\bibliography{citations}

\end{document}